\documentclass[journal,twoside,web]{ieeecolor}
\usepackage{tikz}
\usetikzlibrary{arrows.meta, positioning, calc, backgrounds, fit}%

\usepackage{jsen}
\usepackage{cite}
\usepackage{amsmath,amssymb,amsfonts}
\usepackage{algorithmic}
\usepackage{graphicx}
\usepackage{textcomp}
\usepackage{wrapfig}

\usepackage{bm}
\usepackage[nolist]{acronym}

\usepackage{booktabs}
\usepackage{tabularx}
\usepackage{siunitx}
\DeclareSIUnit{\rad}{rad}
\DeclareSIUnit{\hertz}{Hz}

\newcommand{\bfA}{\bm{A}}

\newcommand{\bfB}{\bm{B}}
\newcommand{\hbfB}{\hat{\bm{B}}}

\newcommand{\bfE}{\bm{E}}

\newcommand{\bfF}{\bm{F}}

\newcommand{\bfk}{\bm{k}}

\newcommand{\bfM}{\bm{M}}

\newcommand{\bfN}{\bm{N}}

\newcommand{\bfp}{\bm{p}}

\newcommand{\hbfr}{\hat{\bm{r}}}

\newcommand{\bfR}{\bm{R}}

\newcommand{\hbfR}{\hat{\bfR}}

\newcommand{\bfs}{\bm{s}}

\newcommand{\bfw}{\bm{w}}

\newcommand{\bfx}{{\bm{x}}}

\newcommand{\hbfx}{\hat{\bm{x}}}

\newcommand{\bfy}{\bm{y}}

\begin{acronym}
\acro{NN}[NN]{Neural Networks}
\end{acronym}

\begin{acronym}
\acro{TL}[TL]{Tolles-Lawson}
\end{acronym}

\begin{acronym}
\acro{GNSS}[GNSS]{Global Navigation Satellite System}
\end{acronym}

\begin{acronym}
\acro{MagNav}[MagNav]{Magnetic Anomaly Navigation}
\end{acronym}

\begin{acronym}
\acro{EKF}[EKF]{Extended Kalman Filter}
\end{acronym}

\begin{acronym}
\acro{RMSE}[RMSE]{Root-Mean-Square Error}
\end{acronym}

\begin{acronym}
\acro{CRLB}[CRLB]{Cramér-Rao Lower Bound}
\end{acronym}

\begin{acronym}
\acro{ML}[ML]{Machine-Learning}
\end{acronym}

\begin{acronym}
\acro{IMU}[IMU]{Inertial Measurement Unit}
\end{acronym}

\begin{acronym}
\acro{ECEF}[ECEF]{Earth-centered Earth-fixed}
\end{acronym}

\begin{acronym}
\acro{FOGM}[FOGM]{First-Order Gauss-Markov}
\end{acronym}

\begin{acronym}
\acro{IGRF}[IGRF]{International Geomagnetic Reference Field}
\end{acronym}

\begin{acronym}
\acro{WGS84}[WGS84]{World Geodetic System 1984}
\end{acronym}

\begin{acronym}
\acro{INS}[INS]{Inertial Navigation System}
\end{acronym}

\begin{acronym}
\acro{RMSE}[RMSE]{Root Mean Square Error}
\end{acronym}

\begin{acronym}
\acro{MC}[MC]{Monte-Carlo}
\end{acronym}

\begin{acronym}
\acro{MSE}[MSE]{Mean Squared Error}
\end{acronym}

\begin{acronym}
    \acro{OPM}[OPM]{optically pumped magnetometer}
\end{acronym}

\begin{acronym}
    \acro{ASD}[ASD]{Amplitude Spectral Density}
\end{acronym}

\begin{acronym}
    \acro{ADEV}[ADEV]{Allan Deviation}
\end{acronym}

\begin{acronym}
    \acro{NV}[NV]{Nitrogen-Vacancy}
\end{acronym}

\begin{acronym}
    \acro{DNV}[DNV]{Diamond Nitrogen-Vacancy} 
\end{acronym}

\begin{acronym}
    \acro{ZFS}[ZFS]{Zero-Field Split}
\end{acronym}

\begin{acronym}
    \acrodef{FIR}[FIR]{Finite Impulse Response}
\end{acronym}
\usepackage{cleveref}
\crefformat{equation}{(#2#1#3)}
\crefrangeformat{equation}{(#3#1#4)-(#5#2#6)}
\crefname{section}{Sec.}{Secs.}
\crefname{figure}{Fig.}{Figs.}
\crefname{table}{Tab.}{Tabs.}
\usepackage[caption=false,font=footnotesize]{subfig}

\def\BibTeX{{\rm B\kern-.05em{\sc i\kern-.025em b}\kern-.08em
    T\kern-.1667em\lower.7ex\hbox{E}\kern-.125emX}}
\definecolor{abstractbg}{rgb}{0.89804,0.94510,0.83137}
\begin{document}

\title{Scalar and Vector Airborne Platform Calibration Using Quantum and Classical Magnetometers and Inertial Sensors}

\author{Antonia~Hager,~%
        Torleiv~H.~Bryne,~%
        and~Mia~Juki\'{c} %
\thanks{Antonia Hager is with Airbus and the Department of Engineering Cybernetics at Norwegian University of Science and Technology (NTNU), Antonia.Hager@airbus.com.}
\thanks{Torleiv~H.~Bryne is with  the Department of Engineering Cybernetics at NTNU, torleiv.h.bryne@ntnu.no}%
\thanks{Mia Juki\'{c} is with the Netherlands Organisation for Applied Scientific Research (TNO), mia.jukic@tno.nl}%
}

\IEEEtitleabstractindextext{%
\fcolorbox{abstractbg}{abstractbg}{%
\begin{minipage}{\textwidth}%
\begin{wrapfigure}[19]{r}{3.5in}%
\includegraphics[width=3.4in]{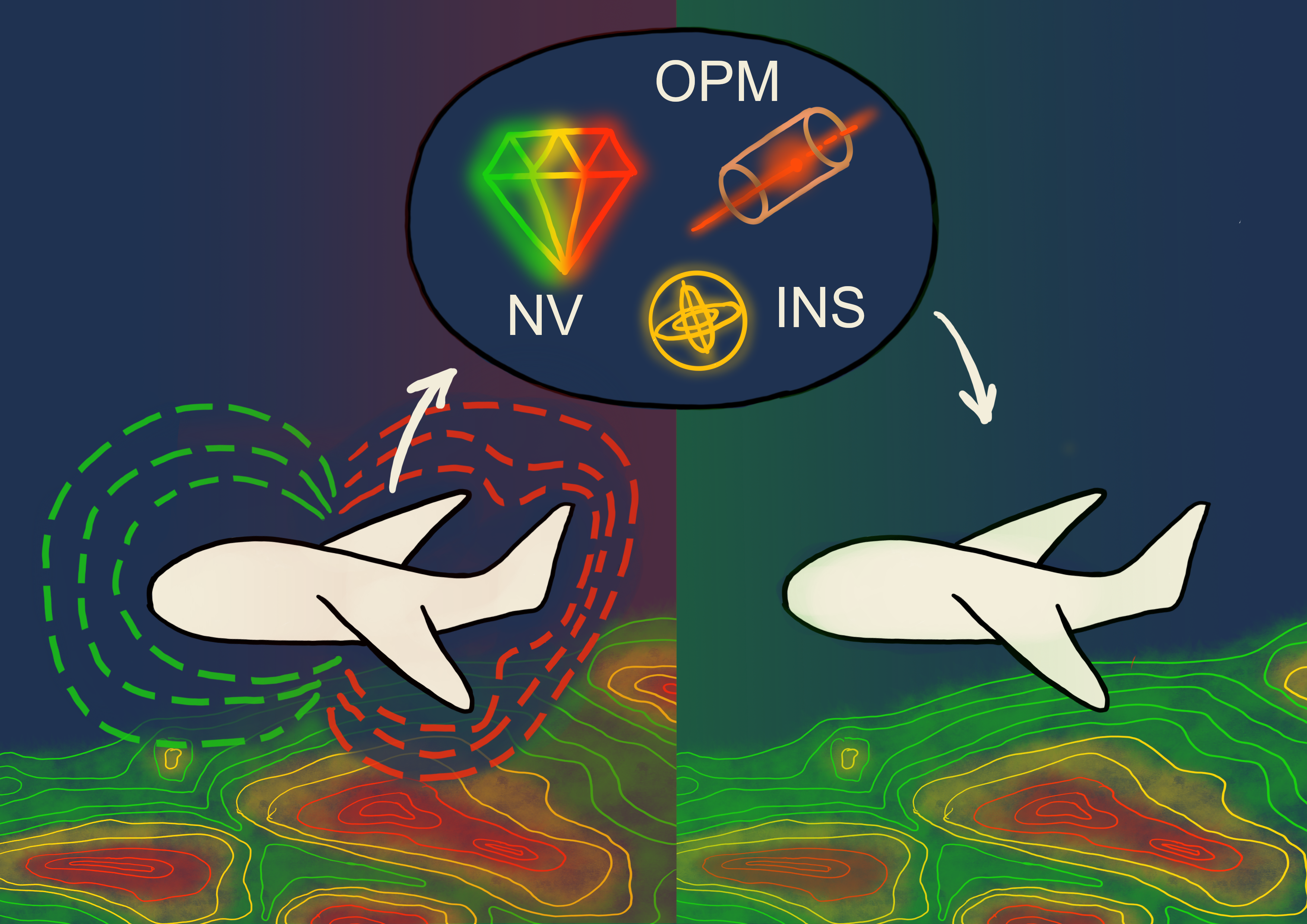}%
\end{wrapfigure}%
\begin{abstract}
Airborne magnetometry requires rigorous calibration to isolate geomagnetic signals from sensor errors and platform magnetic fields. This magnetic compensation is needed for applications like geophysical exploration and magnetic anomaly navigation. The standard approach utilizes a quantum scalar Optically Pumped Magnetometer (OPM) and a less sensitive fluxgate vector sensor for attitude information. This configuration typically results in a scalar approximation of the platform field.

Advancements in high-sensitivity Diamond Nitrogen-Vacancy (NV) vector magnetometers now enable a re-evaluation of the standard hardware configuration and full vector calibration models. We show through rigorous theoretical analysis that scalar calibration models are robust to misalignment. 
Vector calibration models, however, are intrinsically first-order sensitive to attitude errors, irrespective of the accuracy of the magnetic field measurements. These errors arise from inaccurate representation of the background field direction in the body frame, and can amplify small orientation errors into noticeable calibration residuals.

Using realistic sensor models and flight trajectories, we evaluate different sensor configurations for magnetic calibration and assess the use of onboard Inertial Navigation Systems (INS) as an independent attitude reference to enable stable compensation.
Our results suggest that quantum vector magnetometers like NV sensors are not sufficient to solve the attitude bottleneck for airborne vector magnetic calibration. However, as a single sensor capable of providing both absolute field and directional measurements, they may offer benefits regarding sensor placement, synchronization, and scalar calibration accuracy.
\end{abstract}

\end{minipage}}}

\maketitle
\section{Introduction}
\label{sec:introduction}
\IEEEPARstart{M}{agnetic} field mapping is a fundamental component of geophysical exploration and a key requirement for large-scale deployment of \acf{MagNav}, which provides a robust complement to Global Satellite Navigation Systems (GNSS) \cite{Canciani2022, Muradoglu2025}.  Thereby, the correct removal of the aircraft's own magnetic field is a bottleneck for obtaining high-fidelity geomagnetic measurements. Stable, accurate sensors are necessary to separate external signal from onboard interference, but successful calibration also relies on correct knowledge of the external reference field during the calibration procedure.
Incentivized by the maturation of quantum sensors, interest in leveraging the high sensitivity \acf{NV} center-based vector magnetometers for geomagnetic surveying and positioning is increasing. However, despite their promise, it remains an open question whether these next-generation quantum magnetometers deliver practical benefits in airborne magnetometry.

\subsection{Recent advances in Quantum Magnetometry}
\label{sec:intr_quantum_magn}
Quantum magnetometers began to see widespread use in geophysical applications with the invention of \acp{OPM}, highly sensitive sensors that use atomic vapor to measure magnetic fields \cite{tierney2019optically}. A laser first initializes the atomic spins into a dark state, which then precess at a frequency proportional to the magnetic field strength (Larmor frequency). Small changes in the magnetic field change the spin population and thus the absorption of the probe light, enabling femtotesla-level measurements without cryogenics. Beyond geophysical studies, \acp{OPM} are also widely used to detect extremely weak magnetic fields, such as those generated by the human brain \cite{brookes2022magnetoencephalography}.

On the other hand, Nitrogen-Vacancy (NV) Centers in Diamond represent a more recent solid-state approach to quantum magnetometry. An NV center is a defect in the diamond lattice where two neighbouring carbon atoms are replaced by a nitrogen atom and a vacancy. It behaves like a controllable quantum spin. Green laser light initializes the spin, microwaves drive transitions between spin states, and the spin states are read out via fluorescence. Magnetic fields shift the spin resonance (Zeeman effect), and the fixed lattice geometry allows NV-based magnetometers to function as absolute-field or vector sensors with high precision~\cite{Fescenko2020, sturner_integrated_2021, Halde2025, Childress2025, pandey_qdsim_2026}. This feature is one of the biggest advantages of \acf{NV} sensors compared to \acp{OPM}, which can only measure the absolute value of the magnetic field. 
These advances in sensor development are generating renewed interest in quantum technologies for navigation. Recent theoretical work leverages unique sensor capabilities, such as those provided by NV centers, for long-range magnetic gradient matching \cite{le_distributed_2025}, or the availability of precise vector magnetic data to explore trade-offs in vector-based \ac{MagNav} systems \cite{Canciani2020}.
In addition to conventional magnetic field sensing, \acf{NV} magnetometers show promise for other applications as well. For example, they allow nanoscale imaging of electrical current flow in 2D materials such as graphene, revealing microscopic transport features \cite{dockx2025}. Notably, while this work focuses on magnetic quantum sensing, research into quantum inertial sensors --- \ac{NV}-based solid-state gyroscopes being one among several other technologies --- is also on the advance \cite{li_quantum_2026}.
\subsection{Platform Calibration}
\label{ssec:intr_calibration}
Airborne measurements used to produce geomagnetic maps are corrupted by platform-generated magnetic interference that can exceed the signals of interest by orders of magnitude \cite{Canciani2022} and must be compensated for.
To address this challenge, the standard approach for aeromagnetic calibration is the \acf{TL} model \cite{Gnadt2022a, leliak_identification_1961}, which traditionally employs a first-order 1D approximation that projects the platform interference onto the direction of the Earth’s main field.
It leverages the high absolute accuracy of scalar \acf{OPM}, using less stable vector fluxgate sensors only for directional reference. 
To mitigate the impact of sensor errors on calibration, Han et al. \cite{han_modified_2017} showed that the residual interference that is dominated by the fluxgate errors can be improved by including a sensor error model to transform the model coefficients before regression.

The \acf{TL} model accounts for the aircraft’s induced and permanent fields, as well as eddy currents, all of which are referred to as ‘static’ fields \cite{Canciani2022} because the underlying physical phenomena do not change in time, even though the interference magnitude depends on roll, pitch, yaw, and their derivatives.

We note that the \ac{TL} model does not take into account effects from onboard electronic systems, AC and DC currents, temperature-induced nonlinearities, and moving aircraft parts.
Several approaches to improve this standard calibration model were proposed to include nonlinear interference, involving more refined models \cite{jukic2024applications} or data-driven compensation \cite{Gnadt2022b, Moradi2024, Wang2025}. 
In this study, we focus solely on the static contributions captured by the \ac{TL} model, as realistic modeling of time-varying sources is challenging and could be avoided by good sensor placement. We believe that removing the static part of the platform field remains highly relevant, even if it is not sufficient for all types of vehicles.

The emergence of \ac{NV} magnetometers challenges the two-magnetometer solution by potentially enabling a standalone, single-sensor solution. Furthermore, the prospect of precise vector data invites a reconsideration of the full 3D vector calibration model. 
Recent maritime 3D calibration and crustal mapping trials utilizing \ac{NV} vector magnetometers highlighted the challenge of vector calibration due to the need for precise attitude data that is difficult to obtain during vessel maneuvers \cite{frontera_shipboard_2018, fleig_maritime_2018}.
In spaceborne applications, this sensitivity is traditionally alleviated using extremely precise star trackers for attitude reference \cite{liebe_startrackers_1995}. Previous studies have further shown that using an onboard \ac{INS} for attitude reference can improve absolute field calibration \cite{Liu2025}. It remains to be quantified how one- and three-dimensional calibration could benefit from more accurate vector magnetometer measurements, and if three-dimensional calibration is viable on small, dynamic airborne platforms.

\subsection{Main contribution}

In this work, we first theoretically quantify magnetic platform calibration errors introduced by inaccurate sensor measurements. We show the inherent capabilities and limitations of vector and scalar calibration models and how they are impacted by typical sensor errors.

To reflect how these insights translate to actual platform calibration performance, we then simulate calibration scenarios. For this, realistic sensor error models based on the available literature on \acp{OPM}, fluxgate and \ac{NV} magnetometers, and own experiments are used to model platform-corrupted magnetic signals picked up by the sensors. 
The selected sensor models, while only representing a snapshot of current developments and performances in magnetometer technology, fundamentally reflect different qualities of scalar and vector magnetic measurements and thus serve as an illustration of possible results using sensors with certain error characteristics.

\subsection{Organization}
The paper is organized as follows: \cref{sec:tl_basics} recaps the \ac{TL} model and derives the error dependencies and magnitudes when modeling the absolute and three-dimensional platform fields. \cref{sec:sensor_models} and \cref{sec:signal_generation} describe the sensor error models and simulation of onboard measurements, and in \cref{sec:calibration}, calibration results are presented.

\section{Scalar and Vector Tolles-Lawson Calibration}
\label{sec:tl_basics}

\subsection{Notation}

\begin{table}
\centering
\caption{Notation, Reference Frames, and Mathematical Conventions.}
\setlength{\tabcolsep}{3pt}
\begin{tabular}{|p{70pt} p{150pt}|}
\hline
\textbf{Notation} & \textbf{Definition / Convention} \\
\hline
\multicolumn{2}{|l|}{\textit{General Conventions}} \\
$x$ & Scalar quantity (normal-weight) \\
$\bfx$ & Vector quantity (bold lowercase) \\
$\bfA$ & Matrix (bold uppercase) \\
$\hbfx$ & Unit vector (direction) \\
$\tilde{x}, \tilde{\bfx}$ & Approximated or proxy quantity \\
$\dot{\bfx}$ & Time derivative \\
$\left \lVert \cdot \right \rVert $ & 2-norm $\left \lVert \cdot \right \rVert_2 $ \\
\hline
\multicolumn{2}{|l|}{\textit{Reference Frames \& Rotation}} \\
$\{e\}, \{b\}, \{s\}$ & Earth, Body, and Sensor frames \\
$\bfx^b$ & Quantity expressed in frame $\{b\}$ \\
$\bfR_{eb}$ & Rotation matrix from $\{b\}$ to $\{e\}$ \\
\hline
\multicolumn{2}{|l|}{\textit{Magnetic Field \& Calibration}} \\
$\bfB_t, \bfB_e, \bfB_a$ & Total, Earth, and Platform fields \\
$\bfp, \bfN, \bfE$ & Permanent, Induced, and Eddy parameters \\
$\bfA_{\text{TL}}, \bfx_{\text{TL}}$ & Regression matrix and parameter vector \\
$\theta, \phi$ & Angles between field components \\
$\alpha, \delta B$ & Angular (attitude) and magnitude errors \\
\hline
\end{tabular}
\label{tab:notation_definitions}
\end{table}

Notation conventions used here are summarized in \cref{tab:notation_definitions}. An exception is the notation of the magnetic flux density vector $\bfB$ (commonly -- despite being physically incorrect -- also referred to as magnetic field strength), as a capital letter, in line with physics convention.
We assume perfect alignment of the sensor and platform/body frames, $\{s\} \equiv\{b\}$, without loss of generality, since sensor alignment errors are a linear transformation on the calibration parameters.

\subsection{The \acf{TL} Model}
\begin{figure}[tb]
    \centering
        \resizebox{0.5\linewidth}{!}{%
        \begin{tikzpicture}[>=stealth, thick, scale=1.2]
    \coordinate (O) at (0,0);
    \coordinate (Be) at (4,0);
    \coordinate (Bt) at (4.5,1.5);
    
    \draw[->, blue] (O) -- (Be) node[midway, below] {$\bfB_e$};
    \draw[->, red] (Be) -- (Bt) node[midway, right] {$\bfB_a$};
    \draw[->, black] (O) -- (Bt) node[midway, above left] {$\bfB_{t}$};
    
    \draw (3.6,0) arc (180:71.5:0.4);
    \node at (3.4,0.3) {$\theta$};
    
    \draw (0.8,0) arc (0:18.4:0.8);
    \node at (1.0,0.15) {$\phi$};
    
\end{tikzpicture}%
        }%
    \caption{The total field $\bfB_t$ measured by a platform magnetometer is composed of the Earth's field, $\bfB_e$, and the aircraft's own platform field $\bfB_a$ (neglecting space weather effects). The angle $\theta$ between $\bfB_e$ and $\bfB_a$ affects the accuracy of the projection approximation of the 1D calibration model. }
    \label{fig:B_field_vectors}
    \vspace{-0.3cm}
\end{figure}
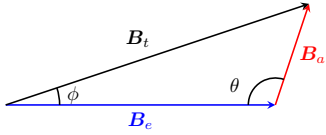

A platform-mounted magnetometer measures a superposition of the Earth's and the platform's magnetic field (\Cref{fig:B_field_vectors}):
\begin{align}
\bfB_t^b &= \bfB_e^b + \bfB_a^b.
\label{eq:Bt_TL}
\end{align}
The classical \ac{TL} model represents the platform field $\bfB_a^b$ as the sum of permanent, induced, and eddy-current fields:
\begin{align}
\bfB_a^b &= \bfp + \bfN \bfB_e^b + \bfE \dot{\bfB}_e^b,
\label{eq:tl_coef_def}
\end{align}
where $\bfp$ is the permanent magnetization vector, $\bfN_{3 \times 3}$ is the induced matrix, and $\bfE_{3 \times 3}$ is the eddy-current matrix. Typically, permanent and induced fields make up the majority of the platform field.
To determine these model coefficients, equation \Cref{eq:Bt_TL} must be solved on a calibration flight data set that contains sufficient excitations of all degrees of freedom. $\bfB_t^b$ is the total field measured in the body frame $\{b\}$ and $\bfB_e^b$ is the external field that generally is a superposition of the Earth's core and crustal field as well as space-weather induced varaiations.
The total field $\bfB_t^b$ on the left-hand side of \cref{eq:Bt_TL} is the field measured by the sensor. The external field $\bfB_e^b$ in the body frame, that appears both as an own term and inside of $\bfB_a^b$, is generally not known and needs to be approximated, replaced or removed from the equation to be able to perform regression on a series of total field measurements.

%

\subsubsection{Scalar Approximation}
Standard \ac{TL} calibration typically employs a first-order scalar approximation of \cref{eq:Bt_TL}, assuming the platform field is sufficiently represented by its projection along the main Earth field direction \cite{Gnadt2022a}:
\begin{align}
B_t \approx B_e + \bfB_a^b \cdot \hbfB_e^b = B_e + B^p_{\text{perm}} + B^p_{\text{ind}} + B^p_{\text{eddy}},
\label{eq:Bt_projection}
\end{align}
where $\cdot$ represents the dot-product. Thus, $ \bfB_a^b \cdot \hbfB_e^b =\bfB_a^{b \top} \hbfB_e^b$. The superscript $\{p\}$ indicates a field component projected along $\hbfB_e^b$, e.g., $B_{\text{perm}}^p = \bfp \cdot \hbfB_e^b$. 
By exploiting the symmetry of $B^p_\text{ind}$, $\bfN$ is reduced to upper triangular. 
Replacing $\bfB_e^b$ and $\dot{\bfB}_e^b$ in \cref{eq:tl_coef_def} by the measured vector $\bfB_t^b$ and its derivative \cite{Canciani2022, Gnadt2022a} yields 
\begin{align}
\begin{split}
B_t &\approx B_e + \bfp \cdot \hbfB_t^b + \bfN \bfB_t^b \cdot \hbfB_t^b + \bfE \dot{\bfB}_t^b \cdot \hbfB_t^b \\ &= B_e + \bfA_{\text{TL}} \bfx_{\text{TL}},
\label{eq:Bt_TL_approx_2}
\end{split}
\end{align}
which can be solved using absolute magnetometer measurements $B_t$ and lower-performance vector measurements $\bfB_t^b$.
An advantage of this scalar form is the applicability of bandpass filtering to the signal $B_t$. By filtering the regression target, one can isolate the frequency bands associated with aircraft maneuvers while removing magnetic gradients of large spatial wavelengths and high-frequency sensor noise, thus removing the remaining $B_e$ from the equation. However, finding the optimal filter requires a trade-off: keeping enough aircraft signal for a robust fit while removing external interference.

While the scalar model is robust, it introduces systematic errors by neglecting components of $\bfB_a^b$ perpendicular to $\bfB_e^b$. 
\subsubsection{Full Vector Calibration}
Solving the vector model \cref{eq:Bt_TL} directly avoids these explicit projection errors but is significantly more sensitive to the precision of the external field reference $\bfB_e$, that needs to be expressed in the platform's body frame.
Furthermore, bandpass filtering is generally not advisable for the vector model, because attitude-induced changes in the magnetic field signals along the sensor's three axes add significant complexity to the frequency spectrum. 

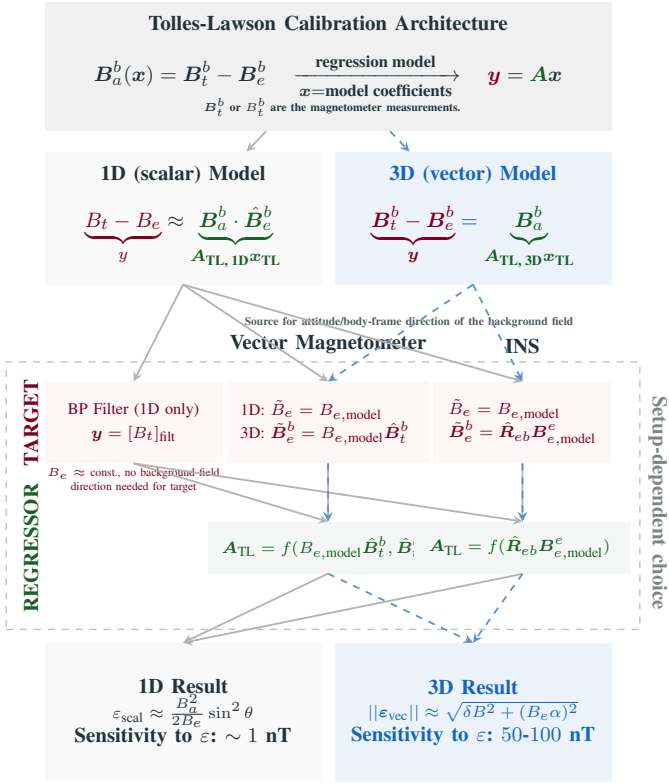
\begin{figure}[tb]
    \centering
    \makebox[\linewidth][c]{%
        \resizebox{\linewidth}{!}{%
        \usetikzlibrary{arrows.meta, positioning, calc, backgrounds, fit}%
\begin{tikzpicture}[
    auto,
    >=stealth,
    /utils/exec={\definecolor{targetcol}{HTML}{800020} 
                  \definecolor{targetbg}{HTML}{FDF7F8}  
                  \definecolor{regcol}{HTML}{1c5e28}
                  \definecolor{regbg}{HTML}{F4F7F5}     
                  \definecolor{structure}{HTML}{263238} 
                  \definecolor{path3d}{HTML}{1565C0}},  
    base_box/.style={rectangle, rounded corners=0pt, align=center, inner sep=6pt, draw=none, font=\small},
    head_box/.style={base_box, fill=gray!12, text=structure, minimum width=8.6cm, font=\small\bfseries},
    m1d_style/.style={base_box, fill=gray!5, text=structure, minimum width=4.2cm},
    m3d_style/.style={base_box, fill=path3d!8, text=path3d, minimum width=4.2cm},
    target_node/.style={base_box, fill=targetbg, text=targetcol, minimum width=2.7cm, minimum height=3.5em},
    regressor_node/.style={base_box, fill=regbg, text=regcol, minimum width=2.7cm},
    path1D/.style={->, thick, draw=gray!60},
    path3D/.style={->, thick, dashed, draw=path3d!70},
]

    \node[head_box] (HEAD) at (0, 13.8) {
        Tolles-Lawson Calibration Architecture \\ 
        \\
        $\bfB_a^b (\bfx) = \bfB_t^b - \bfB_e^b \quad \xrightarrow[\bfx = \text{model coefficients}]{\textbf{regression model}} \quad \textcolor{targetcol}{\bfy} = \textcolor{regcol}{\bfA}\bfx$ \\
        \vspace{0.5pt}
        \tiny $\bfB_t^b$ or $B_t^b$ are the magnetometer measurements.
    };

    \node[m1d_style] (M1D) at (-2.2, 11.5) {
        \textbf{1D (scalar) Model} \\ 
        \\
        $\textcolor{targetcol}{\underbrace{B_t - B_e}_{y}} \approx \textcolor{regcol}{\underbrace{\bfB_a^b \cdot \hbfB_e^b}_{\bfA_{\textbf{TL, 1D}}\bfx_{\textbf{TL}}}}$
    };
    \node[m3d_style] (M3D) at (2.2, 11.5) {
        \textbf{3D (vector) Model} \\ 
        \\
        $\textcolor{targetcol}{\underbrace{\bfB_t^b - \bfB_e^b}_{\bfy}} = \textcolor{regcol}{\underbrace{\bfB_a^b}_{\bfA_{\textbf{TL, 3D}}\bfx_{\textbf{TL}}}}$
    };
    
    \draw[path1D] (HEAD) -- (M1D);
    \draw[path3D] (HEAD) -- (M3D);
    \begin{scope}[yshift=-0.8cm]
    \node[target_node] (T_BPF) at (-2.95, 9.2) {
        \scriptsize BP Filter (1D only)\\ 
        \scriptsize $\textcolor{targetcol}{\bfy = [B_t]_{\text{filt}}}$
    };

    \node[anchor=north, font=\tiny, text=structure!70, inner sep=2pt, align=center, color=targetcol]
          at (T_BPF.south) {$B_e \approx \text{const.}$, no background field \\ direction needed for target};
    
    \node[target_node, align=left] (T_VMAG) at (0, 9.2) {
        \scriptsize 1D: $\textcolor{targetcol}{\tilde{B}_e = B_{e, \text{model}}}$ \\
        \scriptsize 3D: $\textcolor{targetcol}{\tilde{\bfB}_e^b = B_{e, \text{model}} \hbfB^b_t \,\, }$
    };
    \node[target_node, align=left] (T_INS) at (2.95, 9.2) {
        \scriptsize $\textcolor{targetcol}{\tilde{B}_e = B_{e, \text{model}}}$ \\
        \scriptsize $\textcolor{targetcol}{\tilde{\bfB}_e^b = \hbfR_{eb} \bfB_{e, \text{model}}^e}$
    };
    
    \node[rotate=90, color=targetcol, font=\small\bfseries, anchor=base] at (-4.4, 9.2) (T_LABELS) {TARGET};

    \node[regressor_node] (R_VMAG) at (0, 7.3) {
        \scriptsize $\textcolor{regcol}{\bfA_{\text{TL}} = f(B_{e, \text{model}} \hbfB^b_t,\hbfB_t^b)}$
    };
    \node[regressor_node] (R_INS) at (2.95, 7.3) {
        \scriptsize $\textcolor{regcol}{\bfA_{\text{TL}} = f(\hbfR_{eb} \bfB_{e, \text{model}}^e)}$
    };

    \node[rotate=90, color=regcol, font=\small\bfseries, anchor=base] at (-4.4, 7.3) (R_LABELS) {REGRESSOR};

    \begin{scope}[on background layer]
        \node[draw=structure!30, dashed, line width=0.6pt, inner sep=4pt, 
              fit=(T_LABELS) (T_INS) (R_LABELS) (R_INS)] (CHOICEBOX) {};
        
        \node[anchor=south, rotate=-90, font=\small\bfseries, color=structure!60] 
              at (CHOICEBOX.east) [xshift=3pt] {Setup-dependent choice};
    \end{scope}

    \node[font=\small\bfseries\color{structure!80}, anchor=south, yshift=20pt, xshift =-7pt] 
          at ($(T_VMAG.north)!.5!(T_INS.north)$) {\tiny Source for attitude/body-frame direction of the background field};
    \node[font=\small\bfseries\color{structure}, anchor=south] at ($(T_VMAG.north)!(CHOICEBOX.north)!(T_VMAG.north)$) {Vector Magnetometer};
    \node[font=\small\bfseries\color{structure}, anchor=south] at ($(T_INS.north)!(CHOICEBOX.north)!(T_INS.north)$) {INS};

    \node[m1d_style, minimum height=2.1cm] (RES1D) at (-2.2, 4.8) {
        \textbf{1D Result} \\
        \scriptsize $\varepsilon_{\text{scal}} \approx \frac{B_a^2}{2B_e} \sin^2 \theta$ \\
        \textbf{Sensitivity to $\varepsilon$: $\sim 1$~nT}
    };
    
    \node[m3d_style, minimum height=2.1cm] (RES3D) at (2.2, 4.8) {
        \textbf{3D Result} \\
        \scriptsize $||\boldsymbol{\varepsilon}_{\text{vec}}|| \approx \sqrt{\delta B^2 + (B_e \alpha)^2}$ \\
        \textbf{Sensitivity to $\varepsilon$: $50$-$100$~nT}
    };
    
    \end{scope}
    
    \draw[path1D] (M1D.south) -- (T_BPF.north);
    \draw[path1D] (M1D.south) -- (T_VMAG.north);
    \draw[path1D] (M1D.south) -- (T_INS.north); 
    \draw[path3D] (M3D.south) -- (T_VMAG.north);
    \draw[path3D] (M3D.south) -- (T_INS.north);

    \draw[path1D] (T_BPF.south) -- (R_VMAG.north);
    \draw[path1D] (T_BPF.south) -- (R_INS.north);
    
    \draw[path1D] (T_VMAG) -- (R_VMAG);
    \draw[path3D] (T_VMAG) -- (R_VMAG);
    \draw[path1D] (T_INS) -- (R_INS);
    \draw[path3D] (T_INS) -- (R_INS);

    \draw[path1D] (R_VMAG.south) -- (RES1D.north);
    \draw[path1D] (R_INS.south) -- (RES1D.north);
    \draw[path3D] (R_VMAG.south) -- (RES3D.north);
    \draw[path3D] (R_INS.south) -- (RES3D.north);

\end{tikzpicture}%
%
        }%
    }
    \caption{Calibration data sources: Representation of $\bfB_e^b$ in the Target (red) and the Regressor (dark green) based on the attitude source.}
    \label{fig:TL_overview}
    \vspace{-0.5cm}
\end{figure}

An overview of the calibration architecture choices analysed here is given in \cref{fig:TL_overview}. 
For the scalar model, the background field is often simply assumed constant and removed from the target by bandpass filtering.
In the regressor (in all cases), as well as in the target (for 3D calibration or for 1D calibration without bandpass filtering), however, $\bfB_e$ needs to be expressed in the aircraft's body frame by available means. One often-implemented possibility for 1D calibration is to replace $\bfB_e^b$ by $\bfB_t^b$ \cite{Gnadt2022a}. Another alternative is to use a background field model like the IGRF \cite{Alken2021} and express it in the body frame, either with the vector magnetometer as attitude reference, $\tilde{\bfB}_e^b \approx B_{e, \text{model}}^e \hbfB_t^b$, or using an independent attitude source, typically an \ac{INS}, $\tilde{\bfB}_e^b = \hbfR_{eb} \bfB_{e, \text{model}}^e$.

\subsection{Quantification of Fundamental Calibration Error Sources}
The classical \ac{TL} equations represent an idealized physical model that omits several complex dynamics, such as magnetic hysteresis, airframe saturation, and interference from onboard electronics. Beyond these inherent modeling simplifications, the calibration accuracy is fundamentally constrained by geometric approximation errors, reference field uncertainties, and sensor errors.

\subsubsection{Model-specific errors}
Let's assume, first, that we have perfect knowledge of the external background magnetic field in the Earth frame, $\bfB_e^e$ and perfect sensor data from a scalar magnetometer, and from a vector magnetometer or an external attitude reference.
The simplifying assumptions that have to be made to obtain the model parameters using available models and sensor data, but introduce inherent inaccuracies, are:

\begin{enumerate}
    \renewcommand{\theenumi}{(\roman{enumi})}
    \renewcommand{\labelenumi}{\theenumi}
    \item $\left \lVert\bfB_a^b \right \rVert \ll \left \lVert\bfB_e^b\right \rVert$ to obtain the scalar projection model;
    \item $\hbfB_e^b \approx \hbfB_t^b$ in $\bfB_a^b$ (eq. \cref{eq:tl_coef_def}) as proxy of the external field direction and  $\bfB_e^b \approx \left \lVert\bfB_{e, \text{model}}\right \rVert \cdot \hbfB_t^b$ for the external field strength, or
    \item $\bfB_e^b \approx \bfR_{eb}^\top \bfB_{e, \text{model}}^e$, using a model of the external field vector and INS-derived attitude estimates to rotate it to the body frame.
\end{enumerate}
The errors due to these approximations are present even when assuming perfect sensor data. 

\paragraph*{(i)}
The first approximation, $\left \lVert\bfB_a^b\right \rVert \ll \left \lVert\bfB_e^b\right \rVert$, applies at first glance to the scalar model only and can be quantified by looking at higher-order terms of the Taylor expansion of $B_t = \left \lVert \bfB_e + \bfB_a \right \rVert$:
\begin{align}
B_t \approx B_e + B_a \cos \theta + \frac{B_a^2}{2B_e} \sin^2 \theta, \quad \theta = \angle(\bfB_e, \bfB_a).
\end{align}
The first two terms constitute the scalar model, leaving a residual second-order projection error 
\begin{equation}
    \varepsilon_{B_a \ll B_e} \approx \frac{B_a^2}{2B_e} \sin^2 \theta.
\end{equation}
\paragraph*{(ii)}
For the scalar model, replacing $\hbfB_e^b$ by $\hbfB_t^b$ leads to an implicit projection error in $B_a$: Expressing  $\phi = \angle(\bfB_t, \bfB_a)$ as 
\begin{align}
\begin{split}
    \cos(\phi) &= \frac{\bfB_a \cdot \bfB_t}{\left \lVert\bfB_a \right \rVert \left \lVert \bfB_t \right \rVert} 
    = \frac{\bfB_a \cdot(\bfB_a + \bfB_e)}{B_a B_t} \\
    &= \frac{B_a + B_e \cos(\theta)}{B_e + B_a \cos(\theta)}, 
\end{split}
\end{align}
we can then quantify the error as
\begin{align}
\begin{split}
    \varepsilon_{\hbfB_e \approx \hbfB_t} &= \bfB_a^b \cdot \hbfB_e^b - \bfB_a^b \cdot \hbfB_t^b \\
    &= B_a (\cos(\phi) - \cos(\theta))\\
    &= B_a^2 \frac{1 - \cos^2(\theta)}{B_e + B_a \cos(\theta)} = B_a^2 \frac{\sin^2(\theta)}{B_e + B_a \cos(\theta)} \\
    &= \frac{B_a^2}{B_e} \sin^2(\theta).
\end{split}
\label{eq:scalar_error_Be_Bt}
\end{align}
Interestingly, both errors introduced in the scalar model have the same form and order of magnitude. For a \SI{500}{\nano\tesla} platform field in a \SI{50000}{\nano\tesla} background, they would peak at 2.5 and \SI{5}{\nano\tesla} when $\theta = 90^\circ$.
Furthermore, the substitution of $\bfB_t^b$ for $\bfB_e^b$ within the regressor $\bfA_{\text{TL}}$ (Eq. \ref{eq:Bt_TL_approx_2}) introduces a secondary coupling error of order $\mathcal{O}(B_a^2/B_e)$. This error manifests as a slight bias in the induced and eddy-current coefficients, as the model attempts to regress the platform field against a reference already contaminated by platform interference.

For a full three-dimensional calibration of the magnetic vector field using \cref{eq:Bt_TL} for regression, replacing $\hbfB_e^b$ by $\hbfB_t^b$ in the regressor implicitly forces $\bfB_a$ to be parallel to $\bfB_t$,
\begin{align}
    \bfB_a^b = \bfB_t^b - \bfB_{e,\text{model}}^b \cdot \hbfB_t^b \Rightarrow \bfB_{a, \perp}^b \cdot \hbfB_t^b = 0,
\end{align}
leading to a first-order error proportional to the perpendicular component of $\bfB_a^b$,
\begin{align}
    \varepsilon_{\hbfB_e \approx \hbfB_t} &= B_a \sin{\phi} \approx B_a \sin (\theta) \left( 1 - \frac{B_a}{B_e} \cos(\theta)\right) \\
    &\approx B_a \sin(\theta).
\end{align}
Furthermore, using $\bfB_e^b \approx \bfB_t^b$ in the regression target to approximate $\bfB_a^b \approx \tilde{\bfB}_a^b$ also introduces a first-order error 
\begin{align}
    \boldsymbol{\varepsilon}_{\bfB_e^b \approx \bfB_t^b} = \bfB_a^b - \tilde{\bfB}_a^b \approx \bfN \left( \bfB_t^b - \bfB_e^b\right) = -\bfN \bfB_a^b,
\end{align}
neglecting small contributions from the Eddy terms. This effectively reduces the 3D model back to a scalar model while introducing a first-order error from the perpendicular component of the aircraft field.
\paragraph*{(iii)}
 The errors introduced when using an \ac{INS}-based proxy for $\bfB_e^b$ depend on sensor and model errors only.
 
\subsubsection{General effect of a non-perfect $\tilde{\bfB}_e^b \neq \bfB_e^b$}
Fundamentally, all previously listed approximations result in a proxy for the background field in body frame, $\tilde{\bfB}_e^b$, that does not perfectly represent the true $\bfB_e^b$.
Further external error sources that result in a non-perfect $\tilde{\bfB_e} \neq \bfB_e$ are
\begin{itemize}
    \item Background field model errors (both in the core field model and a local map),
    \item Space weather effects: While typically removed from calibration and survey flight measurements using a magnetic ground station, residual errors can remain.
    \item Sensor errors: stochastic and uncorrected scale/alignment errors in the three axes of a vector magnetometer, and heading errors.
\end{itemize}
More general expressions for the error magnitudes that are introduced to the calibration models, depending on the magnitude and angle error in the external field proxy $\tilde{\bfB}_e^b$, can be derived by defining
\begin{align}
\tilde{\bfB}_e^b := (B_e + \delta B) \hbfr
\end{align}
where $\delta B$ is the magnitude error (e.g., due to unmodeled crustal anomalies or background field model errors) and \break $\hbfr \cdot \hbfB_e^b := \cos \alpha$ defines the angular error typically caused by sensor misalignment or INS drift.

For the scalar model, similar to \cref{eq:scalar_error_Be_Bt} we get:
\begin{align}
\begin{split}
    \varepsilon_\text{scal} &= \bfB_a^b \cdot \hbfB_e^b - \bfB_a^b \cdot \hbfr \\
    &= B_a \cos(\theta) - B_a \cos(\theta + \alpha) \\
    & \overset{\alpha \ll 1}{\approx} -B_a ( \alpha \sin (\theta) - \frac{\alpha^2}{2} \cos(\theta)),
\end{split}
\end{align}
where we assumed the worst-case of $\alpha$ and $\theta$ both being in the same plane. 

For the vector model, the compensation error $\boldsymbol{\varepsilon}_\text{vec}$ is the difference between the true and estimated external field vectors:
\begin{align}
\begin{split}
\boldsymbol{\varepsilon}_\text{vec} &= \bfB_e^b - \tilde{\bfB}_e^b = \bfB_e^b - (B_e + \delta B) \hbfr.
\end{split}
\end{align}
To find the magnitude $\left \lVert \boldsymbol{\varepsilon}_\text{vec} \right \rVert$, we apply the law of cosines:
\begin{align}
\begin{split}
\left \lVert\boldsymbol{\varepsilon}_\text{vec}\right \rVert^2 &= B_e^2 + (B_e + \delta B)^2 - 2 B_e (B_e + \delta B) \cos \alpha \\
&\overset{\left \lVert \alpha \right \rVert \ll 1}{\approx} B_e^2 + (B_e + \delta B)^2 - 2 B_e (B_e + \delta B) \left(1 - \frac{\alpha^2}{2}\right) \\
&= \delta B^2 + (B_e^2 + B_e \delta B) \alpha^2.
\end{split}
\end{align}
Neglecting the higher-order term $B_e \delta B \alpha^2$, the final magnitude of the calibration error is:
\begin{align}
\left \lVert\boldsymbol{\varepsilon}_\text{vec}\right \rVert \approx \sqrt{\delta B^2 + (B_e \alpha)^2}.
\end{align}
Because $\alpha$ (in radians) gets directly multiplied by $B_e$, the vector model is sensitive to attitude inaccuracies. A $0.1^\circ$ attitude error in a background of \SI{50000}{\nano\tesla} creates a \SI{87}{\nano\tesla} residual. 

For $B_a = \SI{500}{\nano\tesla}$, the same $0.1^\circ$ error results in a calibration residual of only $\approx \SI{0.87}{\nano\tesla}$ at $\theta = 90^\circ$ when using the scalar model. Magnitude errors $\delta B$, whether constant or heading-dependent, are largely absorbed by the TL coefficients during regression.

As summarized in \cref{tab:error_summary}, the scalar model is remarkably robust. The projection errors rarely exceed \SI{1}{\nano\tesla} due to the small magnitude of $\bfB_a$. In contrast, the vector model's reliance on the absolute orientation of the Earth's background field $\bfB^b_e$ means that minor attitude or reference errors ($0.1^\circ$) translate into large residuals ($\approx 10-\SI{100}{\nano\tesla}$ or more). This explains why the vector model, while theoretically more complete, performs worse than the scalar model without a near-perfect attitude reference.
\begin{table}[tb]
\centering
\caption{Typical calibration error magnitudes with regard to background field errors $\delta B$, $\alpha$ and angle $\theta$ between $\bfB_e$ and $\bfB_a$}
\setlength{\tabcolsep}{3pt}
\label{tab:error_summary}
\begin{tabular}{|p{90pt} p{70pt} l|}
\hline
\textbf{Error Source} & \textbf{Scalar Model ($\varepsilon$)} & \textbf{3D Model ($\left \lVert \boldsymbol{\varepsilon} \right \rVert$)} \\ \hline
Projection/Small Field & $\approx \frac{B_a^2}{2B_e} \sin^2 \theta$ & $0$ (exact) \\
Proxy Field ($\mathbf{B}_e \approx \mathbf{B}_t$) & $\approx \frac{B_a^2}{B_e} \sin^2 \theta$ & $\approx B_a \sin \theta$ \\
Magnitude Error ($\delta B$) & absorbed/$\ll 1$~nT & $\approx \delta B$ \\
Attitude Error ($\alpha$) & $\approx B_a \alpha \sin \theta$ & $\approx B_e \alpha$ \\ \hline
\textbf{Typical Sensitivity} & \text   {$\sim 1$~nT} & \text{$\sim 50$--$100$~nT} \\ \hline
\end{tabular}
\end{table}
\section{Quantum and Classical Sensor Models}
\label{sec:sensor_models}
\begin{figure}[tb]
    \centering
    \includegraphics[width=0.7\linewidth]{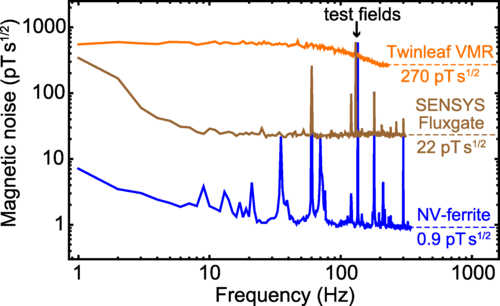}
    \caption{Magnetometer noise characterizations from \cite{Fescenko2020}.}
    \label{fig:fescenko_pic}
\end{figure}

We present realistic sensor models matching typical specifications of state-of-the-art equipment found in the literature. For the standard geomagnetic surveying setup, we model a combination of scalar \ac{OPM}s and fluxgate vector magnetometers. For the emerging \ac{NV} technology, we distinguish between current \textit{lab} and \textit{field} states-of-the-art. It should be noted that no single model can be considered fully representative of its respective technology, as each exists across a broad spectrum of qualities and specifications, with ongoing development continuously pushing performance boundaries. The models presented here are therefore best understood as characteristic points along this spectrum, backed by the literature and chosen to capture a substantial portion of the performance range encountered in practice. An example we base our error model parameterization on is the magnetometer comparison by Fescenko et al. \cite{Fescenko2020} shown in \cref{fig:fescenko_pic}.

\begin{figure}[tb]
    \centering
    \begin{tikzpicture}[
        node distance=1.8cm, 
        auto, 
        >=stealth,
        block/.style={draw, fill=blue!5, rectangle, minimum height=2.5em, minimum width=7.0em, text centered, font=\small, rounded corners=2pt},
        noise/.style={draw, fill=orange!5, rectangle, minimum height=2.5em, minimum width=7.0em, text centered, font=\small, dashed, rounded corners=2pt},
        sum/.style={draw, fill=gray!10, circle, inner sep=2pt},
        label/.style={font=\scriptsize\itshape}
    ]
        \node [block] (physics) {Sensor Physics};
        \node [left of=physics, node distance=2.2cm] (input) {};
        \node [noise, below of=physics, node distance=1.2cm] (stochastic) {Stochastic Noise};
        \node [sum, right of=physics, node distance=2.3cm] (sum) {\small $+$};
        \node [block, right of=sum, node distance=2cm] (lag) {Bandwidth Lag};
        \node [right of=lag, node distance=2.1cm] (output) {};

        \draw [->] (input) -- node[pos=0, anchor=south west, font=\small] {$\bfB_{\text{in}}$} (physics);
        \draw [->] (physics) -- node[pos=0, anchor=south west, font=\small] {$\bfB_{ph}$} (sum);
        \draw [->] (stochastic) -| node[pos=0.6, right, label] {$\eta_{\text{eff}}$} (sum);
        \draw [->] (sum) -- (lag);
        \draw [->] (lag) -- node[pos=1, anchor=south east, font=\small] {$\bfB_{\text{out}}$} (output);

        \node[label, above of=physics, node distance=0.6cm] {Thermal, Heading, Bias};
        \node[label, below of=stochastic, node distance=0.6cm] {$\sigma_{\bfw}, f_{\text{knee}}, \nu$};
        \node[label, above of=lag, node distance=0.6cm] {Low-pass Filter};
    \end{tikzpicture}
    \caption{Error model architecture for simulated magnetometers.}
    \label{fig:sensor_flow}
\end{figure}
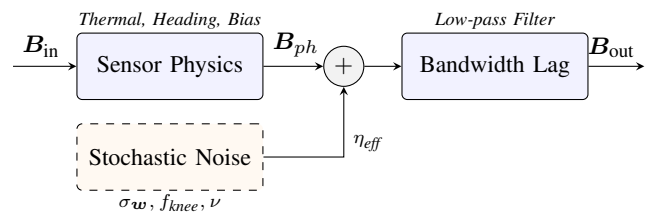

As illustrated in \cref{fig:sensor_flow}, all sensor models share a common error generation pipeline comprising individual physics-based errors, a stochastic noise component $\eta$, and a bandwidth lag.

The stochastic noise is generated using a frequency-domain synthesis method to ensure the output power spectral density matches the physical specifications of each instrument grade. The noise is defined by three parameters: 
\begin{itemize}
    \item the white noise floor $\sigma_w$ (shot-noise limit), 
    \item the corner frequency $f_{knee}$ (the 1/f transition point), and 
    \item the spectral slope $\nu$.
\end{itemize}
An ideal noise power profile in the frequency domain is constructed using
\begin{align}
    S_{target}(f) = \sigma_{\bfw} \left[ \left( \frac{f_{knee}}{\max(f, \epsilon)} \right)^{\nu/2} + 1 \right], \quad \epsilon \ll 1.
    \label{eq:asd_profile}
\end{align}
This profile scales a complex Gaussian white noise vector $\mathbf{Z}(f) \sim \mathcal{CN}(0,1)$ generated in the frequency domain. The final noise time-series $\eta_t$ is obtained via Inverse Fast Fourier Transform (IFFT) and captures the transition from high-frequency white noise to low-frequency drift.

After adding the stochastic noise to the signal that passed through the individual sensor physics, all models incorporate a first-order \ac{FIR} filter to represent the finite response time of the instrumentation:
\begin{align}
B_{out, t} = \beta (B_{ph,t} + \hat{\eta}_t) + (1 - \beta) (B_{ph,t-1} + \hat{\eta}_{t-1})
\label{eq:bandwidth_filter}
\end{align}
where $B_{ph}$ represents the deterministic physics-based field and $\eta_{\text{eff}}$ the geometrically penalized noise. The smoothing factor $\beta = \frac{1}{1 + \frac{f_s}{2\pi f_{BW}}}$ is derived from the sensor's specific bandwidth $f_{BW}$ and the system sampling rate $f_s$.

The Parameters used for the different sensor models are summarized in \cref{tab:sensor_parameters}.
\begin{table}[tb]
\centering
\caption{Error Model Parameters for Modeled Magnetometers}
\label{tab:sensor_parameters}
\begin{tabular}{|l| c c c c c|}
\hline
\textbf{Parameter} & \textbf{Unit} & \textbf{OPM} & \textbf{Fluxgate} & \textbf{NV Field} & \textbf{NV Lab} \\ 
\hline
$\sigma_{\bfw}$     & nT/$\sqrt{\text{Hz}}$ & 0.003 & 0.022 & 0.5   & 0.0009  \\
$f_{knee}$      & Hz                    & 0.5   & 1.0   & 10    & 15     \\
Slope ($\nu$)& -                     & 1.0   & 1.0   & 2.0   & 1.5   \\
$f_{BW}$        & Hz                    & 400   & 60    & 200   & 1000   \\
$\lambda_{\text{t}}$   & \%                    & -    & -   & 0     & 99.5   \\
\hline
\end{tabular}%
\vspace{2pt}
\begin{flushleft}
\scriptsize \textit{Note: \ac{OPM} and fluxgate values are based on QTFM specifications and measurements, and SENSYS Mag-03 benchmarks \cite{Fescenko2020}. NV grades represent a progression from uncompensated field units to optimized lab designs \cite{Fescenko2020, sturner_integrated_2021, Halde2025, Childress2025}.}
\end{flushleft}
\vspace{-0.3cm}
\end{table}

\subsection{Optically Pumped Magnetometer}
Scalar OPMs are the state-of-the-art for absolute magnetic flux density measurements in surveying. While offering high sensitivity, their performance is alignment-dependent. To avoid signal-to-noise ratio degradation or signal loss when the magnetic field is aligned too closely with the sensor axis, some manufacturers build sensors incorporating complementarily aligned gas cells.

\begin{figure}[tb]
    \centering
    \includegraphics[width=0.9\linewidth]{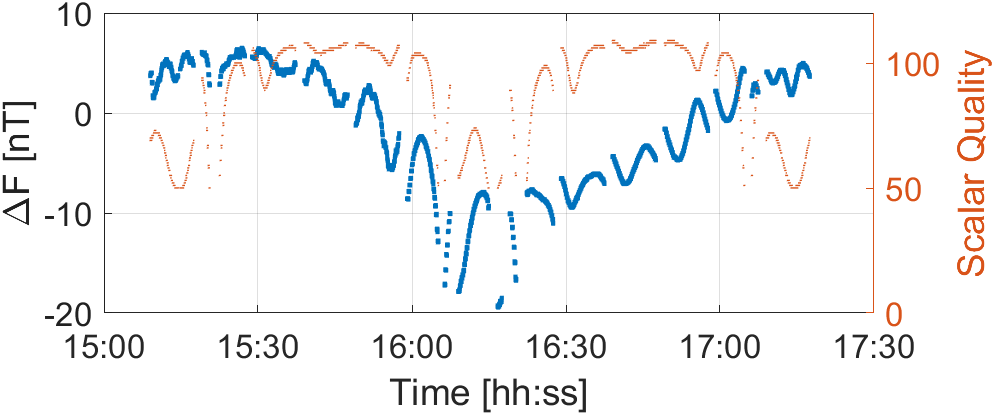}
    \caption{Measured Heading Errors $\Delta \bfF$ in the total field measurements of the QTFM Gen-2 for rotation around two axes. A Gemsys GSM-90 magnetometer was used as reference.}
    \label{fig:quspin_heading_err}
\end{figure}

We observed these typical heading errors in an experiment where the QuSpin QTFM Gen-2 was placed in a magnetically quiet location, only measuring the Earth's background field, and rotated around two of its axes. \cref{fig:quspin_heading_err} shows the signal, corrected for fluctuations in the external field.
Supported by the observed magnitude of the heading errors, we model the deterministic heading error as a multi-harmonic expansion:
\begin{align}
\delta_{h} = \sum_{n \in {1,2,4}} c_n \cos(n \psi)
\label{eq:opm_heading_error}
\end{align}
where $\psi$ is the angle between the external field and the optical axis and the coefficients are set to $c_1 = 0.2$, $c_2 = 2.5$, and $c_3 = 0.5$ nT.

The sensor's signal amplitude degrades as the vector approaches the equatorial dead zone, resulting in an amplified noise floor
\begin{align}
\eta_{\text{eff}} = \eta_{\text{opm}} \frac{1}{\max(\left \lVert\cos(\psi)\right \rVert, \epsilon)},
\end{align}
where the small regularization constant $\epsilon = 0.1$ prevents numerical instability. If $\theta$ falls within a dead zone of half-width $\psi_{dz} = 5^{\circ}$ caused by too close alignment of the total field with the sensors' measurement axis, the model simulates loss-of-lock with high-variance Gaussian noise to emulate the behavior of the control electronics as they are ``searching'' to re-lock on the frequency of the Larmor precession.
The processed scalar value is then passed through a \SI{400}{\hertz} bandwidth filter to match the electronic response of the laser-locking loop.

\subsection{Fluxgate Magnetometer}
The fluxgate model represents high-end commercial instruments (e.g., SENSYS Mag-03, \cite{Fescenko2020}) characterized by the magnetic properties of their Permalloy core.
The geometric error is modeled by multiplying the true signal by an upper triangular matrix $\bfM$ representing the non-orthogonality of the physical coils, and a scaling vector $\bfs$. The temperature-dependent output is calculated as
\begin{align}
\bfB_{ph} = (\bfM \bfB_{in} \circ (1 + \bfk_s \Delta T)) + \bfk_o \Delta T
\end{align}
where $\circ$ is the Hadamard-product, $\Delta T$ the deviation from a reference temperature ($\SI{20}{\degree}C$) and $\bfk_s$ and $\bfk_o$ are linear thermal gain and offset coefficients. The stochastic noise in a fluxgate magnetometer is dominated by Barkhausen noise, modeled as $1/f$ pink noise ($\nu = 1$) with a characteristic floor of $\SI{22}{\pico\tesla/\sqrt{\hertz}}$ \cite[~Fig. 5]{Fescenko2020}. The final output includes a sharp bandwidth cutoff at \SI{60}{\hertz}.

\subsection{\ac{NV} Diamond Magnetometer (NV)}
The \ac{NV} sensor models are categorized into two grades to represent the current technology: \textit{lab} and \textit{field}.
The field grade represents integrated, portable prototypes where sensitivity is currently limited, particularly due to uncompensated thermal drift. The lab grade represents high-performance optical bench setups \cite{Fescenko2020, Graham2024, Childress2025, Halde2025}.

Comprehensive simulation frameworks exist to model the full quantum physics of optical detected magnetic resonance (ODMR) in \ac{NV} diamonds \cite{pandey_qdsim_2026}. We use a more macroscopic model that focuses on the most significant error and noise sources.  
The \ac{NV} sensor model simulates the multi-stage transformation of the ground truth magnetic vector in the sensor frame, $\bfB_{in}$, caused by internal dynamics in the diamond, electronics, and packaging. We model the errors using
\begin{align}
    \bfB_{ph} = (\bfM \bfB_{in} \circ \bfs) +  \frac{(1 - \lambda_{\text{t}})}{\gamma_{\text{e}}} \left( \frac{dD}{dT} \cdot \Delta T \right)\hat{\bm{n}}_{nv}.
\end{align}
Orthogonality and scaling errors from soft-iron distortion and packaging are represented by the upper triangular matrix $\bfM$ and the scaling vector $\bfs$. 
The second term accounts for the dependence of the temperature of \ac{ZFS} along the $[1,1,1]$ crystal axis $\hat{\bm{n}}_{nv}$. $\lambda_{\text{t}}$ is the thermal compensation efficiency, $\frac{dD}{dT} \approx -74.2$ kHz/K is the \ac{ZFS} shift coefficient, and $\gamma_{\text{e}}$ is the gyromagnetic ratio \cite{abe_tutorial_2018}. Note that NV sensors are fundamentally multi-purpose due to the additional dependency of their Hamiltonian on parameters like temperature, strain and electric field \cite{pandey_qdsim_2026}. 
We assume no thermal compensation for the field-grade sensor, and almost-perfect compensations for the lab and future-grade sensors (see \cref{tab:sensor_parameters}).

To account for the \ac{NV} center's asymmetric sensitivity, a geometric penalty is applied to the stochastic noise component $\eta$ similar to that for \ac{OPM}:
\begin{align}
    \eta_{\text{eff}} = \eta_{\text{nv}} \frac{1}{\max(\left \lVert\cos \psi\right \rVert, \epsilon)}.
\end{align}
The constant $\epsilon = 0.15$ was chosen to model the SNR tracking limit before the sampling loop loses fidelity.
The resulting signal is finally processed through the common \ac{FIR} filter \cref{eq:bandwidth_filter} to reflect the sampling limits of the microwave-optical control loop.
\begin{figure}[tb]
  \centering
    \includegraphics[width=\columnwidth]{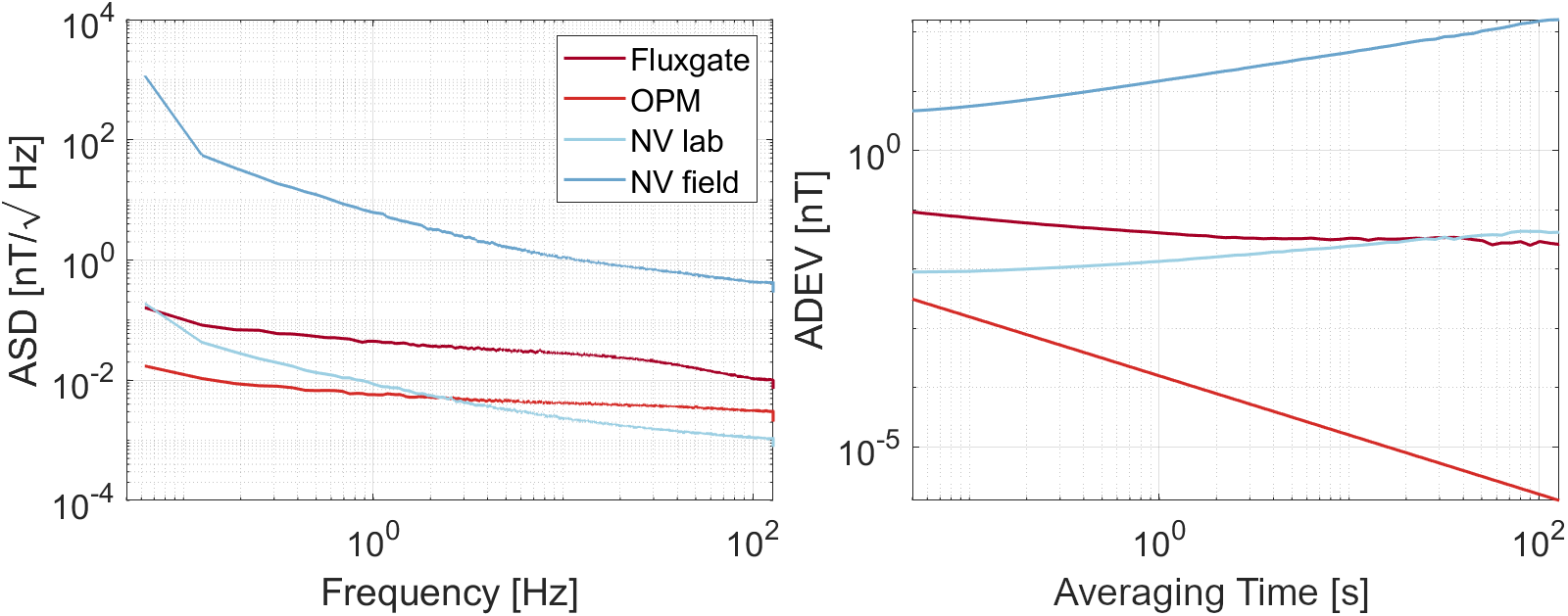}
        \caption{\acf{ASD} and \acf{ADEV} of simulated \SI{256}{Hz} sensors' measurements in a static \SI{50000}{\nano\tesla} external field.}
    \label{fig:asd_adev_const_field}
\end{figure}
\cref{fig:asd_adev_const_field} shows the resulting \acf{ASD} and \acf{ADEV} of the sensor models for a simulated constant-field measurement, meaning no temperature influence or heading-dependent errors. The \ac{ASD} plot visualizes the different stochastic noise levels of the sensor models, while the \ac{ADEV} gives an indication of the stability of the signal. The OPM  excels in both, while the fluxgate has a significantly higher nose level and worse stability. The simulated \ac{NV} sensors, depending on their grade, range from inferior to fluxgates to \ac{OPM}-like characteristics.
\section{Onboard Magnetic Signal Simulation}
\label{sec:signal_generation}
We simulate realistic small aircraft trajectories to evaluate the impact of the sensor errors on aeromagnetic calibration. 

\begin{figure}[tb]
    \centering
    \subfloat[ \label{fig:traj_calibration}]{
        \includegraphics[width=0.6\columnwidth]{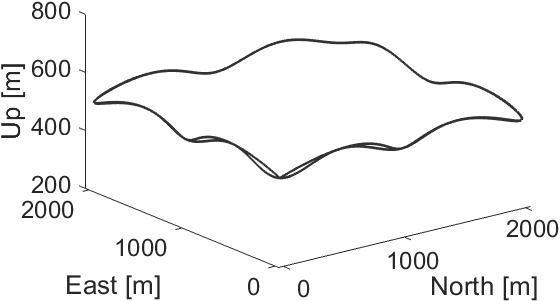}
    }
    
    \subfloat[ \label{fig:traj_validation}]{
        \includegraphics[width=0.6\columnwidth]{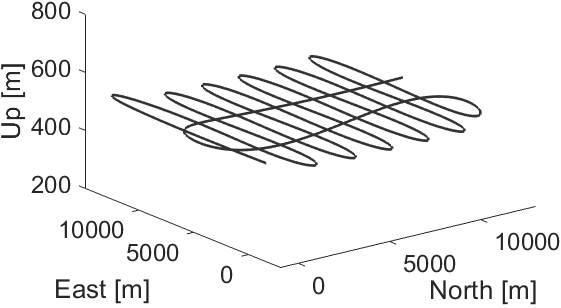}
    }

    \caption{Calibration and Validation flights with average velocities of \SI{60}{\meter\per\second}.}
    \label{fig:trajectory}
\end{figure}

The \textit{calibration} trajectory (\cref{fig:traj_calibration}) is a square that features repeated attitude maneuvers (rolls, pitches, yaws) to excite the platform's magnetic coefficients. The square is flown 5 times, the amplitudes of the approximately 40 consecutive roll and pitch excitations during the \SI{11}{\minute} flight range between [\SI{-55}{\degree}, \SI{15}{\degree}] and [\SI{-10}{\degree}, \SI{10}{\degree}], respectively. For \textit{validation}, we simulate a survey-like flight trajectory with a duration of \SI{58}{\minute} (\cref{fig:traj_validation}).

All data is simulated at \SI{20}{\hertz}. For simplicity, we assume a constant, uniform geomagnetic background field \mbox{$\bfB_e^e = \SI{50000}{\nano\tesla} \cdot \hbfB_e^e$} across the entire flight area. The total clean onboard magnetic vector signal is then calculated as the sum of the external field and the platform interference as defined in \cref{eq:Bt_TL}.
To generate the platform interference $\bfB_a^b$, we use the full \ac{TL} model \cref{eq:tl_coef_def}. $\bfB_e^e$ is rotated into the platform's body frame using the true attitude $\bfR_{eb}$, $\bfB_e^b = \bfR_{eb}^{\top} \bfB_e^e$. 

We simulate two distinct platform field scenarios by choosing coefficients $\bfx_{TL, gt}$:
\begin{itemize}
    \item Random (normalized) coefficients in a range between 10 and 70, resulting in a typical small UAV platform field ranging between 40 and \SI{150}{\nano\tesla} ($\varnothing \, \SI{63}{\nano\tesla}$) \cite{Jukic2024} and an average alignment between $\bfB_e$ and $\bfB_a$, measured as the mean of $\cos (\theta)$ over the calibration flight dataset, of $0.27$.
    \item Perpendicular stress test: strong y-component in $\bfB_a^b$, resulting in a platform field ranging between 500 and \SI{1100}{\nano\tesla} ($\varnothing \, \SI{700}{\nano\tesla}$) and an average alignment between $\bfB_e$ and $\bfB_a$ of $0.03$ to challenge the projection approximations.
\end{itemize}
Sensor mounting errors can be neglected because any static misalignment is mathematically absorbed into the estimated \ac{TL} coefficients during the calibration process.

\begin{figure}[tb]
  \centering
    \includegraphics[width=\columnwidth]{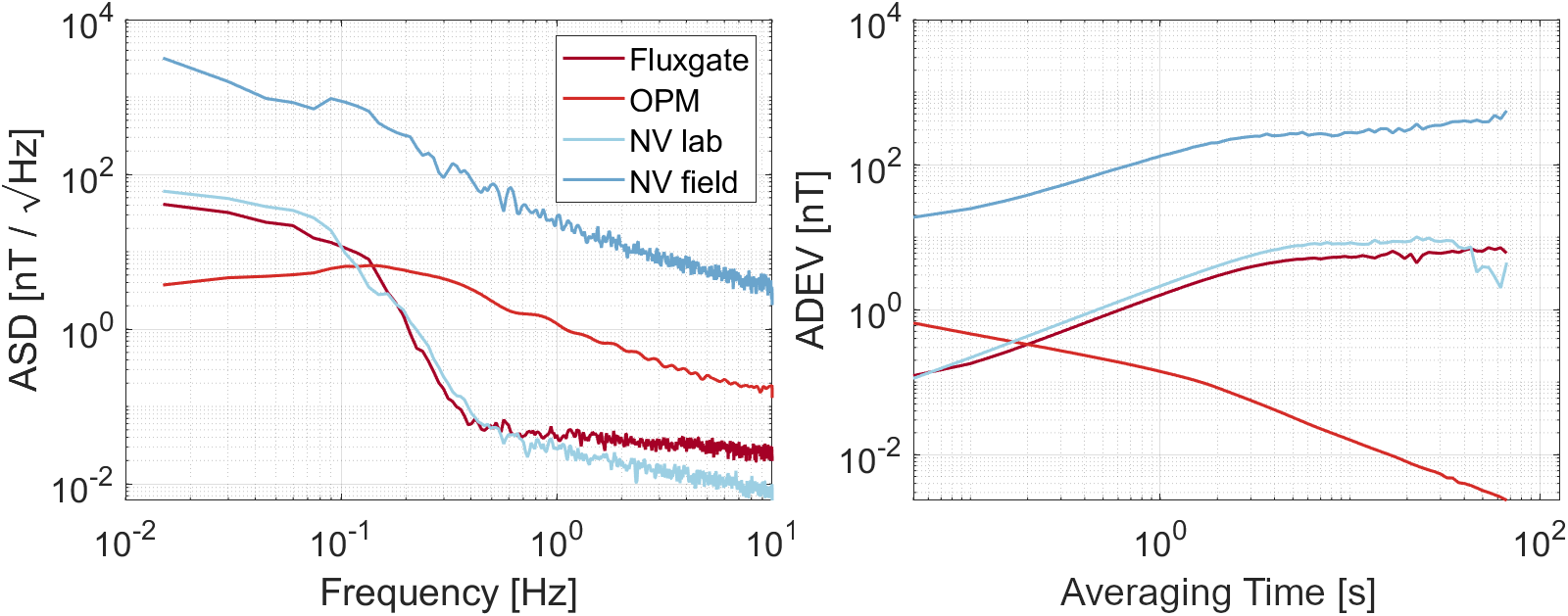}
    \caption{ASD and ADEV of the simulated 20 Hz onboard magnetometer measurements that include altitude-dependent temperature drifts.}
    \label{fig:asd_adev_trajectories}
    \vspace{-0.4cm}
\end{figure}

Lastly, the simulated scalar and vector magnetometer data time series are corrupted using the noise models described in \cref{sec:sensor_models}.
To account for environmental effects, we model the sensors' internal temperature $T_s$ based on altitude $h$ and a first-order thermal lag ($\tau = \SI{300}{\second}$). This temperature profile drives the deterministic \ac{ZFS} and bias drifts in our \ac{NV} and fluxgate models, respectively. Comparing the noise statistics of the simulated in-flight data \cref{fig:asd_adev_trajectories} to the static-field measurements \cref{fig:asd_adev_const_field}, in particular, the low-frequency noise is amplified for all sensors because of their heading-dependent noise characteristics, and the stability of all sensors but the \ac{OPM} exceeds single-digit nanotesla deviations due to temperature drifts.

We then use the measurements generated along our test calibration trajectory to obtain the \ac{TL} coefficients via linear regression using both the projection model and the full 3D model.

\section{Results}
\label{sec:calibration}
We perform regression on our calibration trajectory without applying any bandpass filtering to obtain the calibration model coefficients $\hbfx_{\text{TL}}$, and validate the results by applying the resulting calibration parameterization to platform-interference corrupted magnetic measurements along the validation flight path. The time derivatives of the magnetic field are obtained by first-order numerical differentiation after applying an optional lowpass filter with a cutoff frequency of 1 Hz to suppress the impact of high-frequency noise.
When using a background field model, we assume it to have no errors,  $\bfB_e^e = \bfB^e_{e,\text{model}}$.
As summarized in \cref{fig:TL_overview}, we evaluate different sensor combinations as regression target and regressors.

The calibration performance is measured by calculating the residuals 
\begin{align}
\delta B_e = \left \lVert\tilde{\bfB}_{\text{e}}^b - \bfB^b_{\text{e}, gt}\right \rVert = \left \lVert(\tilde{\bfB}^b_{\text{t}}-\bfA_{TL}\hbfx_{TL}) - \bfB^b_{\text{e}, gt}\right \rVert.
\end{align}
Note that the residual has a lower bound defined by the sensor noise floor, $\delta B \approx \frac{\sigma_m}{\sqrt{Hz}}$.
\vspace{-0.2cm}

\subsection{Calibration with Ideal Magnetometers}
\label{ssec:cal_perf}
\begin{figure}[tb]
    \centering
    \includegraphics[width=0.8\columnwidth]{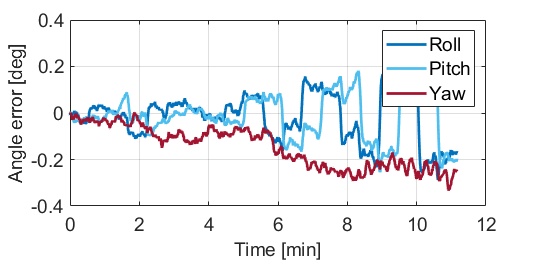}
    \caption{Reference attitude drift of our simulated tactical-grade IMU measurements during the calibration flight range between $\pm \SI{0.3}{\degree}$.}
    \label{fig:attitude_errors}
    \vspace{-0.4cm}
\end{figure}

The accuracy and sensitivity to sensor errors of the 1D and 3D regression models are compared by performing a series of calibrations and validations on data where no errors were applied to the magnetometer measurements. 
We analyze three scenarios:
\begin{enumerate}
    \item Ideal regression target and regressor: We assume perfect knowledge of the external field in body frame (perfect attitude knowledge and no model errors).
    \item Vector magnetometer: We approximate $\bfB_e^b = B_{e,\text{model}} \hbfB_t^b$, where $\bfB_t^b$ is the vector magnetometer measurement.
    \item Independent attitude reference: we approximate $\bfB_e^b = \hbfR_{eb}^\top  \bfB^e_{e,\text{model}}$, where $\hbfR_{eb}$ is the rotation from Earth to body frame, using tactical-grade \ac{INS}-based attitude.
\end{enumerate}
Tactical-grade inertial sensors, as typically mounted on a geo-survey platform, exhibit a drifting attitude error. Our gyroscope errors is characterized by an angular random walk of $3.6 \times 10^{-5}$ rad/$\sqrt{\text{s}}$ and a bias with a standard deviation of $4.8 \times 10^{-6}$ rad/s and $\SI{3600}{\second}$ time constant. An example of the resulting drift is shown for our calibration trajectory in \cref{fig:attitude_errors}. 
\begin{figure*}[!htb]
    \centering
    \includegraphics[width=\linewidth]{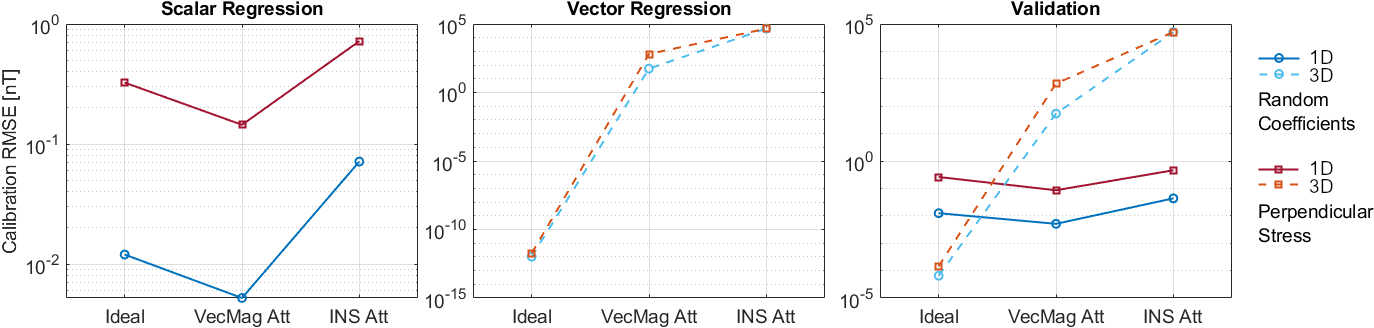}
    \caption{Calibration results and validation with noise-free magnetometer data for the 1D and 3D models.}
    \label{fig:perfect_sensors_results}
\end{figure*}

The calibration performance on the two sets of platforms is summarized \cref{fig:perfect_sensors_results}.  Scalar regression converges reliably using both vector-magnetometer and tactical-grade INS-derived attitude for the regression target. The 3D vector calibration model is successful only in the theoretical, ideal case of perfect knowledge of the external field in the platform frame. It can be clearly seen from this ideal scenario that, if it were possible to obtain perfect knowledge of $\bfB_e^b$, the vector model would yield extreme accuracy. However, as soon as transitioning to realistic assumptions, where $\bfB_e^b$ has to be derived from vector magnetometer or \ac{INS} data, attitude inaccuracies lead to significant errors in the 3D model as the large Earth-field vector is incorrectly projected, causing field components to leak into orthogonal axes, while the 1D model robustly performs. Despite an observable performance degradation when the platform field has strong perpendicular components, it still achieves sub-nanotesla accuracy.
\vspace{-0.2cm}

\subsection{Calibration with Non-Ideal Magnetometers}
To inspect the effect of typical sensor error characteristics of existing magnetometers, we compare two distinct hardware configurations:
\begin{enumerate}
    \item Fluxgate + OPM: State-of-the-Art setup, distinct sensors for the absolute field and the field direction
    \item \ac{NV} magnetometers: one sensor for both field magnitude and direction, in a lab-grade and a field-grade configuration.
\end{enumerate}

\begin{figure*}[tb]
    \centering
    \includegraphics[width=\linewidth]{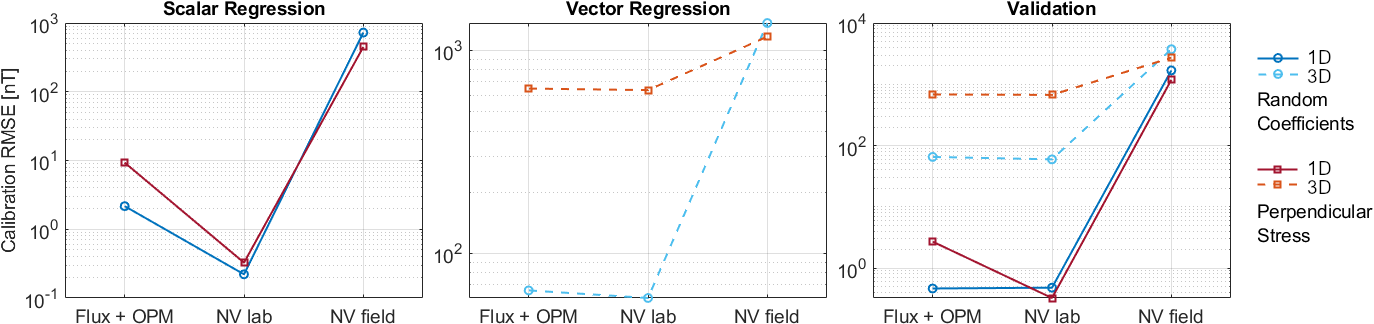}
    \caption{Calibration results and validation across different magnetometers for 1D and 3D models, using vector magnetometer data for $\bfB_e^b$.}
    \label{fig:sensor_error_MAG_results}
            \vspace{-0.5cm}
\end{figure*}

In \cref{fig:sensor_error_MAG_results}, the calibration performance is shown using the respective vector magnetometer measurements to obtain external-field measurements.
Even with small errors in the external reference, the 3D model's accuracy is extremely degraded, and the well-established 1D model is clearly superior. 
While in our analysis, the state-of-the art sensor configuration employing a fluxgate magnetometer and an \ac{OPM} is visibly superior to using data from our \ac{NV}-field-grade sensor and achieves single-digit accuracy even on the challenging platform field with strong perpendicular components, the lab-grade \ac{NV} configuration leads to even lower, sub-nanotesla, average calibration errors.
It should be noted again that the achievable accuracy is fundamentally limited by the noise floor of the sensor (and, if present, the model errors).

Low calibration errors are confirmed when the calibration model obtained is applied to the validation trajectory, interestingly leading to even smaller errors in the resulting estimate $\tilde{\bfB_e}$. This might be due to a much less extreme attitude maneuvering along the validation trajectory.

\begin{figure*}[!htb]
    \centering
    \includegraphics[width=\linewidth]{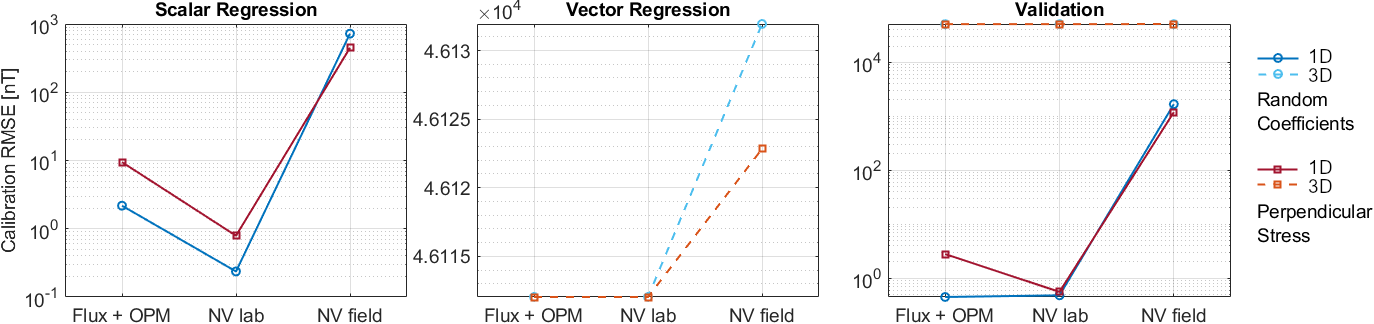}
    \caption{Calibration results and validation for both 1D and 3D models across different magnetometer setups using tactical-grade INS-derived attitude for $\bfB_e^b$.}
    \label{fig:INS_perfect_vs_drifting_results}
        \vspace{-0.5cm}
\end{figure*}

Finally, we want to explore the effects of using \ac{INS}-derived attitude on calibration as an alternative to a vector magnetometer. For that, we use $\bfB_e^b = \hbfR_{eb}^\top \bfB^e_{e,\text{model}}$ for both the regression target and the regressor, as we did for the perfect magnetometer scenarios in \cref{ssec:cal_perf}.
In \cref{fig:INS_perfect_vs_drifting_results} we show the calibration performance when using this drifting attitude reference to rotate a background field model to the body frame.

A slight improvement over using magnetometer-based attitude is observed for the 1D model using the lab-grade NV sensor, while the 3D model remains limited by the attitude-induced error leakage.
This good performance of the 1D model with a drifting attitude reference leads to the question of whether a better external attitude reference can improve the calibration even further. To verify this, we ran the same calibration tests again with a perfect attitude reference representing an accurate attitude obtainable from e.g., GNSS aided-\ac{INS} or with even higher grade \acp{INS}. No significant improvement was achieved, which leads to the conclusion that magnetometer errors in the scalar measurement still dominate the calibration error.
\section{Discussion}
The simulation results provide a picture of the trade-offs between established scalar-based calibration and the potential of high-accuracy vector magnetometers.

The difficulties and limitations of a 3-dimensional platform calibration are not a mathematical limitation of the model itself, but rather a consequence of the lack of accurate knowledge of the background field in the body frame due to background field model and sensor errors. Even though the vector model provides extreme accuracy in the ``ideal" scenario (Section \ref{ssec:cal_perf}), when transitioning to realistic assumptions, the small errors introduced by imperfect attitude knowledge and/or model inaccuracies lead to significant leakage of vector components between the axes and continue to render vector calibration extremely challenging. Thus, this work supports and explains the results of \cite{frontera_shipboard_2018}.

The performance of the lab-grade \ac{NV} configuration in our simulation shows the potential of accurate vector sensors in aeromagnetics. By consolidating both high-precision scalar magnitude and vector orientation into a single sensor, \ac{NV} magnetometers have the potential to eliminate the mechanical misalignment and synchronization issues of the traditional Fluxgate + OPM two-sensor setup.

In our stress-test scenario, the lab-grade \ac{NV} configuration achieved a factor-two improvement over the state-of-the-art setup, reducing calibration residuals from $\sim \SI{1}{\nano\tesla}$ to roughly $0.4\,\text{nT}$. This suggests that with higher accuracy vector measurements available in the field, the bottleneck for 3D calibration will shift from sensor noise to attitude precision.

We also showed the stability of the 1D model when using significantly drifting tactical-grade \ac{INS} attitude data. Being less sensitive to the aforementioned ``attitude leakage'', calibration residuals are comparable to a magnetometer-only calibration, supporting the findings of Liu et al. showing \ac{INS}-based calibration with attitude errors in the range of $\pm \SI{1}{\degree}$ \cite{Liu2025}.

The potential of using high-accuracy \ac{INS} attitude as external attitude reference can be evaluated comparing the ideal scenario in \cref{fig:perfect_sensors_results} to the results with realistic magnetometer errors in \cref{fig:INS_perfect_vs_drifting_results}, While, theoretically, the ideal case shows the potential of high-grade \ac{INS} when the magnetometers are effectively error-free, in the realistic case an improvement could only be seen using the 1D regression for the lab-grade \ac{NV} sensor (from \SI{0.8}{\nano\tesla} down to \SI{0.4}{\nano\tesla} when perfect \ac{INS} data was used. The challenging 3D calibration fails in our simulation due to the magnetometer errors and cannot be recovered with a high-grade \ac{INS} attitude.

While we were separating the platform calibration from the sensor errors to quantify the impacts of uncompensated sensor errors, calibration in reality often is an integrated process without a defined boundary between compensation for the platform interference and the sensor noise. Although we quantify calibration accuracy limits using the \ac{TL} model, the fundamental mechanisms identified apply similarly to a broader scope of magnetic calibration models.

For airborne \ac{MagNav}, \SI{1}{\nano\tesla} is a practical floor for the after-calibration accuracy requirement, which can be reached with state-of-the-art setups. Calibration for online navigation is, as current research suggests, likely to employ adaptive algorithms, including real-time feedback from the system state to handle dynamic aircraft magnetization and sensor errors \cite{Muradoglu2025, hager2026}.
However, the ground-, air- and spaceborne acquisition of the maps and models needed can benefit from advanced sensors \cite{Halde2025, beerden2024}.
While geomagnetic map-based navigation mainly uses the absolute field, a consideration is to use vector magnetometer measurements directly for navigation. Although vector \ac{MagNav} is not that sensitive to magnetometer attitude errors \cite{Canciani2020} since the navigation filter feedback ``pulls'' the vector back into its correct alignment, vector map availability, and 3D platform calibration will still be a challenge \cite{Canciani2020}. If the 3D platform field is not calibrated to near-nanotesla levels, the navigation filter will be unable to distinguish between aircraft maneuvers and genuine crustal gradients.
Better magnetometers --- both absolute and vector type --- directly benefit geomagnetic research and exploitation, or other fields like single-molecule, high-resolution scanning probe sensing or biomedical applications, where accuracies well below single-digit nanotesla are desirable \cite{abe_tutorial_2018, Glenn2017, Stolz2024, Zhang2021, dockx2025}.

Work on the ruggedization of quantum magnetometers is ongoing.
\acp{OPM} are currently prevalently deployed in geophysical applications due to their relative maturity in integrated systems, \ac{NV} magnetometers have yet to demonstrate comparable stability and drift performance outside laboratory conditions. Bias-field-free operation \cite{Childress2025} offers a new path towards reducing the deterministic drifts that are particularly challenging in field-grade configurations.
\section{Conclusion}
This work investigated the performance of classical and quantum magnetometers in scalar 1D and vector 3D airborne platform magnetic calibration by quantifying impact of error sources theoretically and using a simulation framework designed to represent realistic sensor, aircraft attitude, and model approximation uncertainties. Under idealized simulated conditions, where the external magnetic field in the platform's body frame is assumed to be perfectly known, the 3D calibration model achieves near-perfect accuracy. However, when realistic error sources are introduced in the simulations, even small inaccuracies in attitude knowledge or background field representation lead to pronounced leakage of the dominant Earth field component into orthogonal axes, resulting in a severe degradation of vector calibration performance.

Across all simulated scenarios, including idealized and noisy magnetometers as well as \ac{INS}-derived attitude references with representative drift, the established 1D projection model consistently demonstrated robust behavior and minimized calibration residuals. In contrast, the 3D model remained highly sensitive to these simulated imperfections and did not outperform the 1D approach outside of the strictly idealized case. These results provide a simulation-based explanation for the persistent difficulty of achieving reliable vector-based platform calibration.
The simulations further indicate the potential of high-accuracy vector instruments to improve absolute platform field estimation. In particular, the modeled lab-grade \ac{NV} configuration outperformed the simulated Fluxgate + \ac{OPM} reference by combining scalar and vector information within a single sensor. In practice, this can allow for measurements at a single measurement point, eliminating intra-sensor axis misalignment and synchronization errors.

Nevertheless, the results show that calibration accuracy remains fundamentally limited by model constraints, and that improvements in attitude accuracy alone, by employing accurate vector magnetometers or inertial sensors, do not significantly enhance calibration performance within the simulated framework.
Overall, this simulation study confirms the 1D projection model as the most robust and reliable approach for platform magnetic calibration under realistic error assumptions, while clarifying the fundamental limitations that currently restrict the applicability of 3D calibration. 

Future work should extend the presented simulation framework to include additional error sources, such as time-varying background field disturbances and nonlinear interferences, and should validate the conclusions in practice through dedicated calibration and validation flights comparing various sensor setups.

\section*{Acknowledgment}

The authors would like to thank Jan Oechsle for conducting experiments with the QuSpin QTFM Magnetometer in the Brorfelde Observatory, and Nils Olsen for analyzing the data.
The authors would further like to thank Jasper Krauser, Nils Olsen, Tor Arne Johansen, Reinier Tan, Alexander ter Haar, and Fedde van der Meer for carefully reading the manuscript and valuable comments that improved the quality of this paper.
We acknowledge the use of Google Gemini in editing text and code for this publication.

Mia Jukić acknowledges the Netherlands Ministry of Economic Affairs (SMO programme P104) for supporting the underlying research presented in this paper. She also acknowledges the Netherlands Ministry of Defence (research programme V2104) for contributing to the broader knowledge base on sensor technologies.
\bibliographystyle{ieeetr}
\bibliography{refs.bib}

\end{document}